\documentclass[aps,pre,twocolumn,groupedaddress,longbibliography]{revtex4-1}
\usepackage{amsmath,amssymb,graphicx,dsfont,tikz,subfigure,mathtools,color,float}
\usepackage[english]{babel}
\usepackage[utf8]{inputenc}
\usepackage{hyperref}
\usepackage{lipsum}
\usepackage{graphicx}

\newcommand{\br}{\mathbf{r}}

\begin{document}
\title{Approximating microswimmer dynamics by active Brownian motion:\\ Energetics and efficiency}

\author{Jannik Ehrich}
\email{jannik.ehrich@uni-oldenburg.de}
\author{Marcel Kahlen}
\affiliation{Universit\"at Oldenburg, Institut f\"ur Physik, 26111 Oldenburg, Germany}
\date{\today}

\begin{abstract}
We consider the dynamics of a microswimmer and show that they can be approximated by active Brownian motion. The swimmer is modeled by coupled overdamped Langevin equations with periodic driving. We compare the energy dissipation of the real swimmer to that of the active Brownian motion model, finding that the latter can  massively underestimate the complete dissipation. This discrepancy is related to the inability to infer the full dissipation from partial observation of the complete system. We introduce an efficiency that measures how much of the dissipated energy is spent on forward propulsion.
\end{abstract}

\maketitle

\section{Introduction}\label{sec:intro}
Microswimmers are small-scale biological or artificial objects with an active self-propulsion mechanism~\cite{Elgeti2015, Bechinger2016}. 

Their hydrodynamics have been the object of a long-standing interest dating back to Purcell and the famous \emph{Scallop theorem}~\cite{Purcell1977}. Since then, a number of microswimmer models have been introduced, e.g., assemblies of coupled spherical particles which achieve directed motion through their interactions. These include the three-sphere-swimmer by Najafi and Golestanian~\cite{Najafi2004}, of which there has also been an experimental realization~\cite{Leoni2009}, and other similar models~\cite{Avron2005,Felderhof2006,Felderhof2014}.

While microswimmers can have a rather complex structure, their movement is often described by active Brownian motion, i.e., Brownian motion in two or three dimensions with a constant force whose direction undergoes free diffusion.

Although active (e.g., Janus) particles are correctly modeled by active Brownian motion, for microswimmers this approximation is valid at most for the body of the swimmer. This is because it neglects the motion of those degrees of freedom needed to propel it forward. This fact is especially relevant when considering energy dissipation.

In the following, we use stochastic thermodynamics~\cite{Jarzynski2011,Seifert2012} to describe the energetics of small-scale systems. It enables assigning heat and work~\cite{Sekimoto1998} as well as an entropy production~\cite{Seifert2005} to individual trajectories described by overdamped Langevin dynamics and thus provides a framework for analyzing dissipation of stochastic systems. 

It is well known that the presence of hidden slow degrees of freedom has an impact on central results of stochastic thermodynamics~\cite{Rahav2007,Mehl2012,Esposito2012,Bo2014,Uhl2018,Kahlen2018}. Typically, an effective description of the visible degrees of freedom is obtained by employing a coarse-graining scheme. However, the average dissipation inferred from such a description is underestimated~\cite{Esposito2012,Bo2014}. With a concrete model, one is able to quantify the difference between the coarse-grained and the complete dissipation.

Recently, there have been efforts to formulate stochastic thermodynamics for active matter systems~\cite{Ganguly2013,Szamel2014,Chaudhuri2014,Speck2016,Mandal2017,Pietzonka2018}. The discussion revolves around assigning an adequate trajectory-dependent entropy production to the dynamics of active Brownian particles.

However, since active Brownian motion neglects relevant degrees of freedom of the complete microswimmer dynamics, it is interesting to compare the energy dissipation of the approximate description to that of a more complex swimmer model.

Therefore, the aims of this paper are the following: (1) Propose a microswimmer model that consists of two driven coupled colloidal particles and is able to generate self propulsion. (2) Specify how active Brownian motion results from a coarse-graining scheme applied to the model to be able to compare the energy dissipation rates. (3) Having established that active Brownian motion is an approximate process, contrast its dissipation rate with that of the real swimmer and define a swimming efficiency.

\section{Model}\label{sec:model}
The propulsion mechanism of our microswimmer model shall mirror a nonreciprocal periodic shape transformation. A viable approximation of such a swimmer consists of many coupled spherical particles~\cite{Felderhof2006,Felderhof2014} which interact through time-dependent internal forces, yielding the desired shape transformation.

Therefore, we study the most simplified version of this setting: two spherical Brownian particles submersed in a solution at temperature $T$. We assume overdamped dynamics. The particles have different time-dependent mobilities $\nu_1(t)$ and $\nu_2(t)$, respectively, and are coupled by a time-dependent interaction potential $V\left(r;l(t)\right)$ with $l(t)$ controlling the equilibrium separation between the particles. Here, $r$ denotes the distance between the particles at positions $\br_1$ and $\br_2$, respectively.

\begin{figure}[ht]
 \includegraphics[width=\columnwidth]{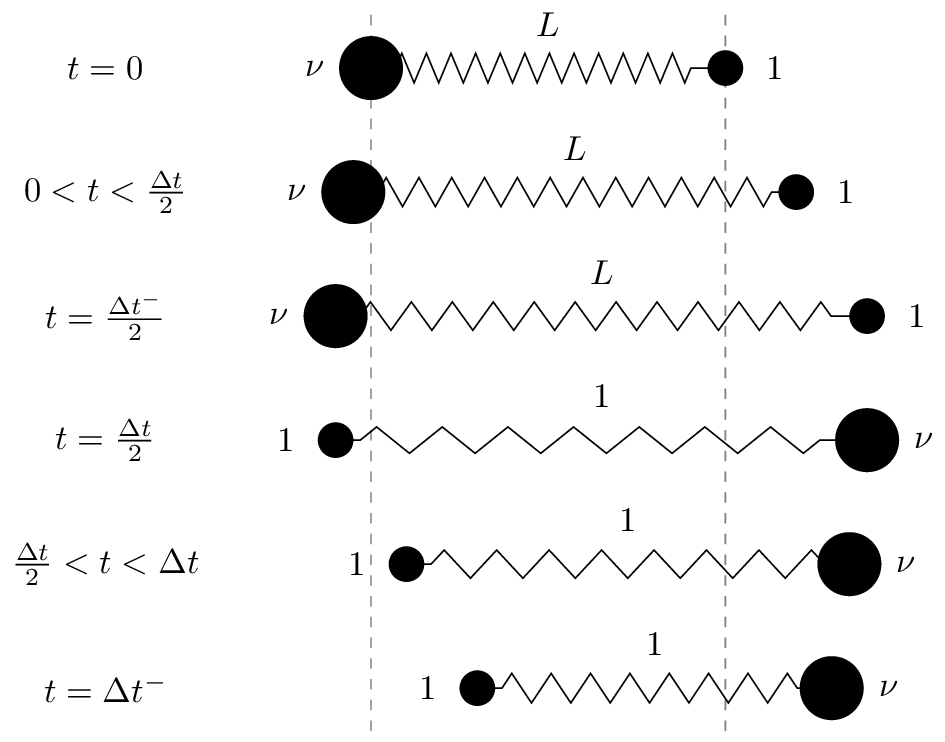}
 \caption{Schematic representation of the microswimmer dynamics. The equilibrium distance of the interaction potential is periodically switched between a long length $L$ and a short length $1$. The individual mobilities are switched between a high mobility $1$ and a low mobility $\nu$ in phase with the length variation.}
 \label{fig_schematicProcess}
\end{figure}

Swimming is achieved by periodically switching the equilibrium distance between a short and a long value and additionally varying the two mobilities between a high and a low value. We choose dimensionless quantities such that the short length and the high mobility are both equal to one. Additionally, we set the Boltzmann constant to unity throughout. The protocol is then given by
\begin{subequations}
\begin{align}
        l(t) &=  \begin{cases} 
               L, & 0 \leq \hspace{-9pt} \mod{(t, \Delta t)} < \frac{\Delta t}{2} \\
               1, & \frac{\Delta t}{2} \leq \hspace{-9pt} \mod{(t, \Delta t)} < \Delta t
                       \end{cases},\\
        \nu_1(t) &=  \begin{cases} 
               \nu, & 0 \leq \hspace{-9pt} \mod{(t, \Delta t)} < \frac{\Delta t}{2} \\
               1, &  \frac{\Delta t}{2} \leq \hspace{-9pt} \mod{(t, \Delta t)} < \Delta t
                       \end{cases},\\
        \nu_2(t) &=  \begin{cases} 
               1, &  0 \leq \hspace{-9pt}\mod{(t, \Delta t)} < \frac{\Delta t}{2} \\
               \nu, & \frac{\Delta t}{2} \leq \hspace{-9pt} \mod{(t, \Delta t)} < \Delta t
                       \end{cases},
\end{align}
\end{subequations}
where $L>1$ is the longer length, $0\leq \nu < 1$ is the lower mobility, and $\Delta t$ is the cycle time. Varying the mobilities can be thought of as inflating or deflating the spheres, which changes the coefficient of Stokes's friction. Figure~\ref{fig_schematicProcess} shows a schematic representation of the swimmer's movement. We also compiled a video illustrating the swimmer's motion in two dimensions~\cite{movie}.

A version of this model has been introduced by Avron \emph{et al}.~\cite{Avron2005}, who also analyzed its hydrodynamics. Here, we incorporate thermal fluctuations and model the dynamics using overdamped Langevin equations
\begin{subequations}\label{eqn_Langevin}
\begin{align}
\dot{\br}_1 &= - \nu_1(t)\,\nabla_1 V(r,l(t)) + \sqrt{2 \nu_1(t) T} \,\boldsymbol{\xi}_1(t), \\
\dot{\br}_2 &= - \nu_2(t)\,\nabla_2 V(r,l(t)) + \sqrt{2 \nu_2(t) T} \,\boldsymbol{\xi}_2(t),
\end{align} 
\end{subequations}
where $\boldsymbol{\xi}_1(t)$ and $\boldsymbol{\xi}_2(t)$ are zero-mean Gaussian white noise terms whose Cartesian components $k$ and $l$ satisfy $\left\langle\xi_i^{(k)}(t)\, \xi_j^{(l)}(t') \right\rangle=\delta_{ij}\,\delta_{kl}\, \delta(t-t')$.

The swimmer's dynamics are reminiscent of a \emph{flashing ratchet}~\cite{Reimann2002}. Here, directed motion is a result of the damping which violates momentum conservation. A similar model implementing a kind of \emph{feedback ratchet} has been introduced by Amb\'{i}a and H\'{i}jar~\cite{Ambia2016,Ambia2017}.

In the following, we will analyze the model first in one and later in two dimensions and show that the center of mass performs active Brownian motion in the limit of small cycle times $\Delta t$.

\section{one-dimensional swimmer}\label{sec:1dswimmer}
For the one-dimensional swimmer we choose a harmonic coupling $V(r; l(t)) = \frac{1}{2} \left(r - l(t)\right)^2$. The particles are at positions $x_1$ and $x_2$, respectively. Their distance is given by $r = x_2-x_1$. The Langevin Eqs.~\eqref{eqn_Langevin} then read
\begin{subequations}\label{eqn_1dLangevin}
\begin{align}
	\dot{x}_1 &= \nu_1 V' + \sqrt{2 \nu_1 T} \xi_1 \\
	\dot{x}_2 &= - \nu_2 V' + \sqrt{2 \nu_2 T} \xi_2 ,
\end{align}
\end{subequations}
where we used $V' := \partial_r V(r; l)$ and dropped the explicit time-dependence. Switching to center of mass $X := \frac{1}{2}(x_1 + x_2)$ and relative coordinates, one obtains
\begin{subequations}\label{eqn_1dRelativeLangevin}
\begin{align}
	\dot{r} &= - (\nu_1 + \nu_2) V' - \sqrt{2 \nu_1 T} \xi_1 + \sqrt{2 \nu_2 T} \xi_2, \\
	\dot{X} &= \frac{\nu_1 - \nu_2}{2} V' + \sqrt{\frac{\nu_1 T}{2}} \xi_1 + \sqrt{\frac{\nu_2 T}{2}} \xi_2 .
\end{align}
\end{subequations}
The ensemble distribution $p(r,X;t)$ evolves according to the corresponding Fokker-Planck equation:
\begin{align}\label{eqn_1dFPE}
	\partial_t\, p(r, X; t) = \mathcal{L}(t)\, p(r, X; t),
\end{align}
with the generator
\begin{align}\label{eqn_1dgenerator}
	\mathcal{L}&(t) := (\nu_1 + \nu_2) \partial_r V' - \frac{\nu_1 - \nu_2}{2} V' \partial_X  \\
													&- T(\nu_1 - \nu_2) \partial_r \partial_X + T(\nu_1 + \nu_2)\partial_r^2 + T \frac{\nu_1 + \nu_2}{4} \partial_X^2. \nonumber
\end{align}

Due to the linear drift and piecewise constant diffusion coefficients in Eq.~\eqref{eqn_1dFPE}, a Gaussian ansatz yields the following evolution equations for the cumulants:
\begin{subequations}\label{eqn_1d_cumulants}
\begin{align}
	\dot{\mu}_r &= -(\nu_1 + \nu_2) (\mu_r - l), \\
	\dot{\mu}_X &= \frac{\nu_1 - \nu_2}{2} (\mu_r - l),\\
	\dot{c}_{rr} &= -2(\nu_1 + \nu_2) c_{rr} + 2 T (\nu_1 + \nu_2), \\
	 \dot{c}_{rX} &= \frac{\nu_1 - \nu_2}{2} c_{rr} - (\nu_1 + \nu_2) c_{rX} - T (\nu_1 - \nu_2),\\
	\dot{c}_{XX} &= (\nu_1 - \nu_2) c_{rX} +  T \frac{\nu_1 + \nu_2}{2} .
\end{align}
\end{subequations}

\begin{figure*}[ht]
\includegraphics[width=0.5\linewidth]{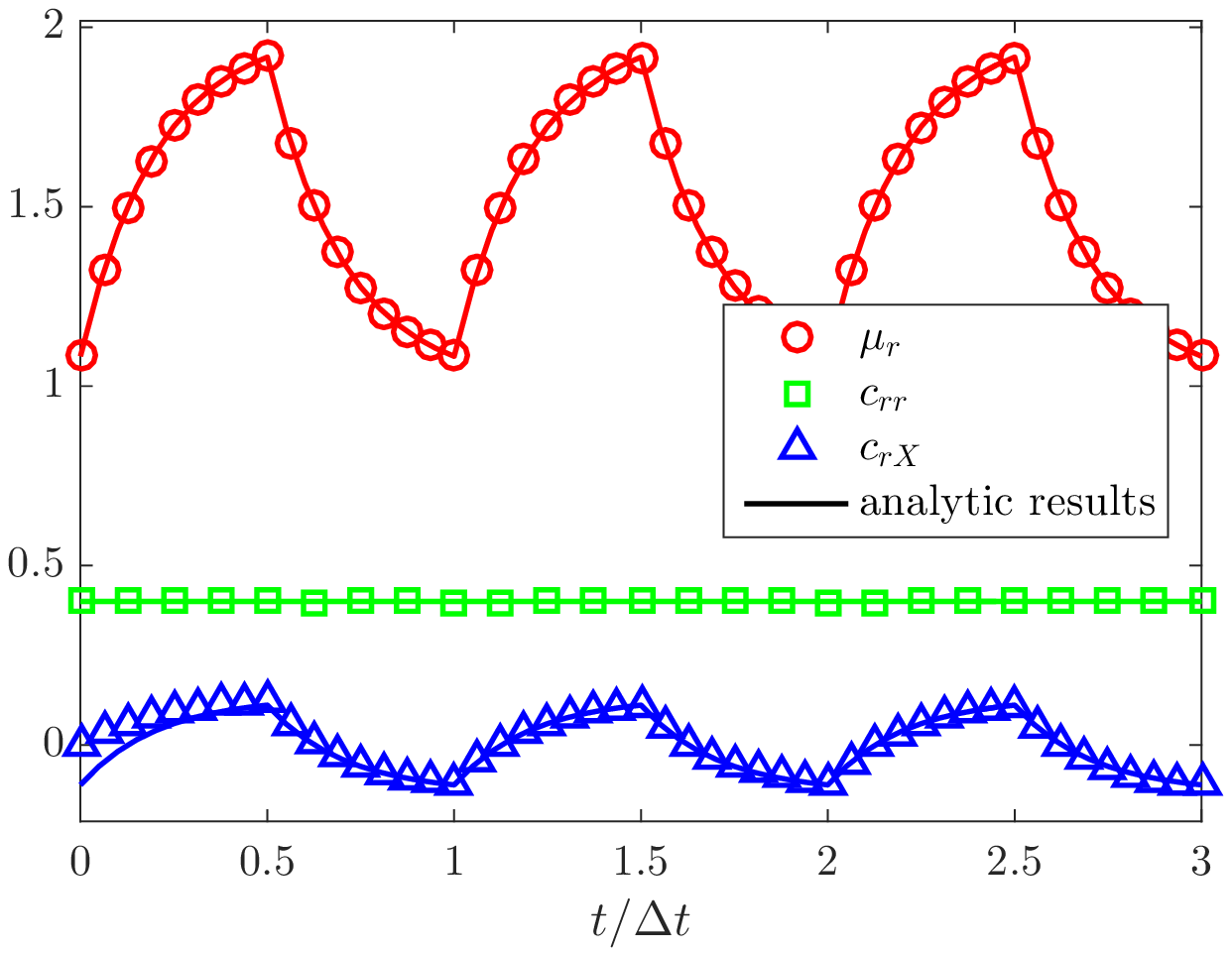}%
\includegraphics[width=0.5\linewidth]{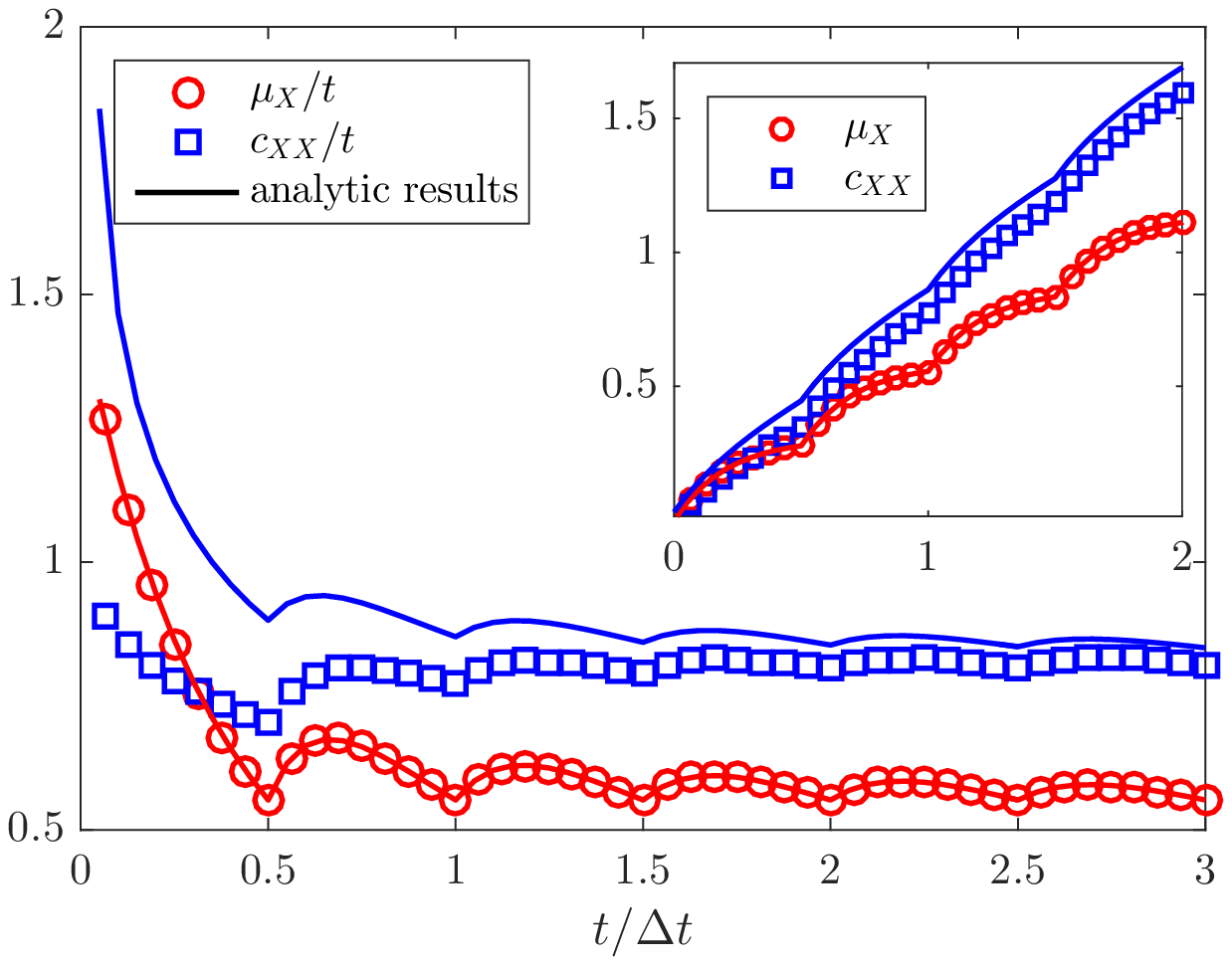}
\caption{Time evolution of the cumulants for the one-dimensional swimmer. The cumulants involving the relative coordinate $r$ (left) are in a periodic steady state. The mean and variance of the center of mass coordinate (right) grow by constant increments during a full cycle. Solid lines show the analytic solutions for system parameters $L=2$, $\nu=0.2$, $T=0.4$, and $\Delta t = 4$. Symbols represent simulations results for $N=10^5$ trajectories with time step $dt = 10^{-3}$. The initial condition of the simulations represents an experimentally realizable situation: The microswimmer is held fixed at $X=0$ and the relative coordinate is allowed to equilibrate. Therefore all trajectories are started from $X=0$ and $r$ is drawn from the periodic steady state which explains the transient relaxation of $c_{rX}$ and $c_{XX}$.}
\label{fig_anaSol1d}
\end{figure*}

Because of the periodic driving, $p(r,X;t)$ does not become stationary. However, the cumulants involving the $r$-coordinate reach a periodic stationary state specified by
\begin{align}\label{eq:eqn_1d_stationaryregime1}
	\mu_r(t + \Delta t) &= \mu_r(t), \nonumber \\
	c_{rr}(t+\Delta t) &= c_{rr}(t), \nonumber \\
    c_{rX}(t+\Delta t) &= c_{rX}(t).
\end{align}
During a full cycle, the remaining cumulants grow by the constant increments $\Delta \mu_X$ and $\Delta c_{XX}$, respectively,
\begin{align}\label{eqn_1d_stationaryregime2}
	\mu_X(t+\Delta t) &= \mu_X(t)+\Delta \mu_X , \nonumber \\  
	c_{XX}(t+\Delta t) &=  c_{XX}(t) + \Delta c_{XX} .
\end{align}

Assuming that the swimmer starts in the periodic stationary regime specified by Eqs.~\eqref{eq:eqn_1d_stationaryregime1}~and~\eqref{eqn_1d_stationaryregime2}, we solve Eqs.~\eqref{eqn_1d_cumulants} with the additional assumption $\mu_X(0) = 0$. The solutions for the mean values in the interval $t \in [0, \Delta t]$ are then given by
\begin{subequations}
\begin{align}
	\mu_r(t) &=  \begin{cases} 
               		L + \frac{\sigma^{-2t}\left(1-L\right)}{1+\sigma^{-\Delta t}}, & \hspace{-1pt} 0 \leq t \leq \frac{\Delta t}{2} \\
               		1 + \frac{\sigma^{-2t+\Delta t}\left(L-1\right)}{1+\sigma^{-\Delta t}}, & \hspace{-1pt} \frac{\Delta t}{2} \leq t \leq \Delta t
                \end{cases}, \label{eqn_1d_solmur} \\
      \mu_X(t) &=  \begin{cases} 
               		\frac{(1-\nu)(1-L)\left( \sigma^{-2t}-1 \right)}{2(1+\nu)\left(1+ \sigma^{-\Delta t} \right)}, & \hspace{-4pt} 0 \leq t \leq \frac{\Delta t}{2} \\
               		\frac{(1-\nu)(1-L)\left( \sigma^{-2t+\Delta t} + \sigma^{-\Delta t}  -2 \right)}{2(1+\nu)\left(1+ \sigma^{-\Delta t} \right)}, & \hspace{-4pt} \frac{\Delta t}{2} \leq t \leq \Delta t
                \end{cases},      
\end{align}
\end{subequations}
where $\sigma := \exp{\left(\frac{1+\nu}{2}\right)}$. Similarly, we obtain $c_{rr}(t) \equiv T$, $c_{rX}(t)$, and $c_{XX}(t)$. We omit the full time dependence of the latter two in favor of brevity. The constant increments are given by
\begin{subequations}
\begin{align}
\Delta \mu_X &= (L-1) \frac{1-\nu}{1+\nu} \tanh\left( \frac{\nu+1}{4} \Delta t \right)\\
\Delta c_{XX} &=\frac{2\nu T}{\nu+1} \Delta t + 2 T \frac{(1 - \nu)^2}{(\nu+1)^2} \tanh\left( \frac{\nu+1}{4} \Delta t \right) .
\end{align}
\end{subequations}

With these results, the full solution can be assembled. It is shown for a representative set of parameters in Fig.~\ref{fig_anaSol1d} together with results from numerical simulations of the Langevin Eqs.~\eqref{eqn_1dLangevin}.

\subsection{Coarse-graining in the limit of short cycle times}\label{sec:1dCG}
Due to the constant increments of the mean and variance of the center of mass coordinate $X$, a measurement of the center of mass position with low time resolution will yield biased diffusion. Indeed, in realistic scenarios tracking of a microswimmer will focus only on the center position. The swimmer's additional degrees of freedom which accomplish propulsion will mostly be too small and too fast to be accurately resolved. Hence, we analyze the model in the limit of very small cycle times $\Delta t \rightarrow 0$ and subsequently integrate out the $r$-variable.

The generator [Eq.~\eqref{eqn_1dgenerator}] is periodic and time independent within each of the two phases. Thus, it may be written as
\begin{align}
	\mathcal{L}(t) = \begin{cases} 
               			\mathcal{L}_1, & 0 \leq \hspace{-9pt} \mod{(t, \Delta t)} < \frac{\Delta t}{2} \\
               			\mathcal{L}_2, &  \frac{\Delta t}{2} \leq \hspace{-9pt} \mod{(t, \Delta t)} < \Delta t
                     \end{cases},
\end{align}
with time-independent generators $\mathcal{L}_1$ and $\mathcal{L}_2$ for the first and the second phases, respectively. For small $\Delta t$, the solution of the Fokker-Planck Eq.~\eqref{eqn_1dFPE} can be expanded up to terms of order $\Delta t$:
\begin{subequations}
\begin{align}
	p\left(r, X; \frac{\Delta t}{2} \right) &= p\left(r, X; 0 \right) + \frac{\Delta t}{2}\, \mathcal{L}_1 \,p\left(r, X; 0 \right), \\
	p\left(r, X, \Delta t \right) &= p\left(r, X; \frac{\Delta t}{2} \right) + \frac{\Delta t}{2} \, \mathcal{L}_2 \, p\left(r, X; \frac{\Delta t}{2} \right).
\end{align}
\end{subequations}
Therefore, 
\begin{align}
	\frac{p\left(r, X; \Delta t \right)-p\left(r, X; 0 \right)}{\Delta t} = \frac{\mathcal{L}_1 + \mathcal{L}_2}{2} p\left(r, X; 0 \right) ,
\end{align}
and for $\Delta t \rightarrow 0$ we obtain the Fokker-Planck equation:
\begin{align}\label{eqn_effectiveFPE}
	\partial_t\, p(r, X; t) = \bar{\mathcal{L}}\, p(r, X; t),
\end{align}
with the effective generator 
\begin{align}\label{eqn_effectiveGenerator}
	\bar{\mathcal{L}} &:= \frac{\mathcal{L}_1 + \mathcal{L}_2}{2},\\
	&= \left(\nu + 1\right) \partial_r \left(r - \frac{L+1}{2}\right) - (L-1)\frac{1-\nu}{4} \partial_X  \nonumber \\
	&\qquad\qquad\qquad\qquad + (1+\nu) T \partial_r^2 + \frac{1+\nu}{4} T\, \partial_X^2,
\end{align}
where we used Eq.~\eqref{eqn_1dgenerator}.

Upon integration of Eq.~\eqref{eqn_effectiveFPE} over $r$, we obtain an effective equation for the center of mass:
\begin{align}\label{eqn_effectiveFPE_1d}
	\partial_t p(X; t) = - \nu_{\mathrm{eff}} f_{\mathrm{eff}}\, \partial_X  p(X; t) + \nu_{\mathrm{eff}}\, T\, \partial_X^2 p(X, t),
\end{align}
with the effective mobility 
\begin{align}\label{eqn_effMobility}
\nu_{\mathrm{eff}} = \frac{1+\nu}{4}
\end{align}
and the constant force 
\begin{align}\label{eqn_effForce}
f_{\mathrm{eff}} = (L-1) \frac{1-\nu}{1+\nu}.
\end{align}
Note that, as expected, the constant force vanishes in the limits $\nu \rightarrow 1$ (no change of mobilities) and $L \rightarrow 1$ (no change of the equilibrium distance).

\begin{figure}[ht]
\includegraphics[width=1\linewidth]{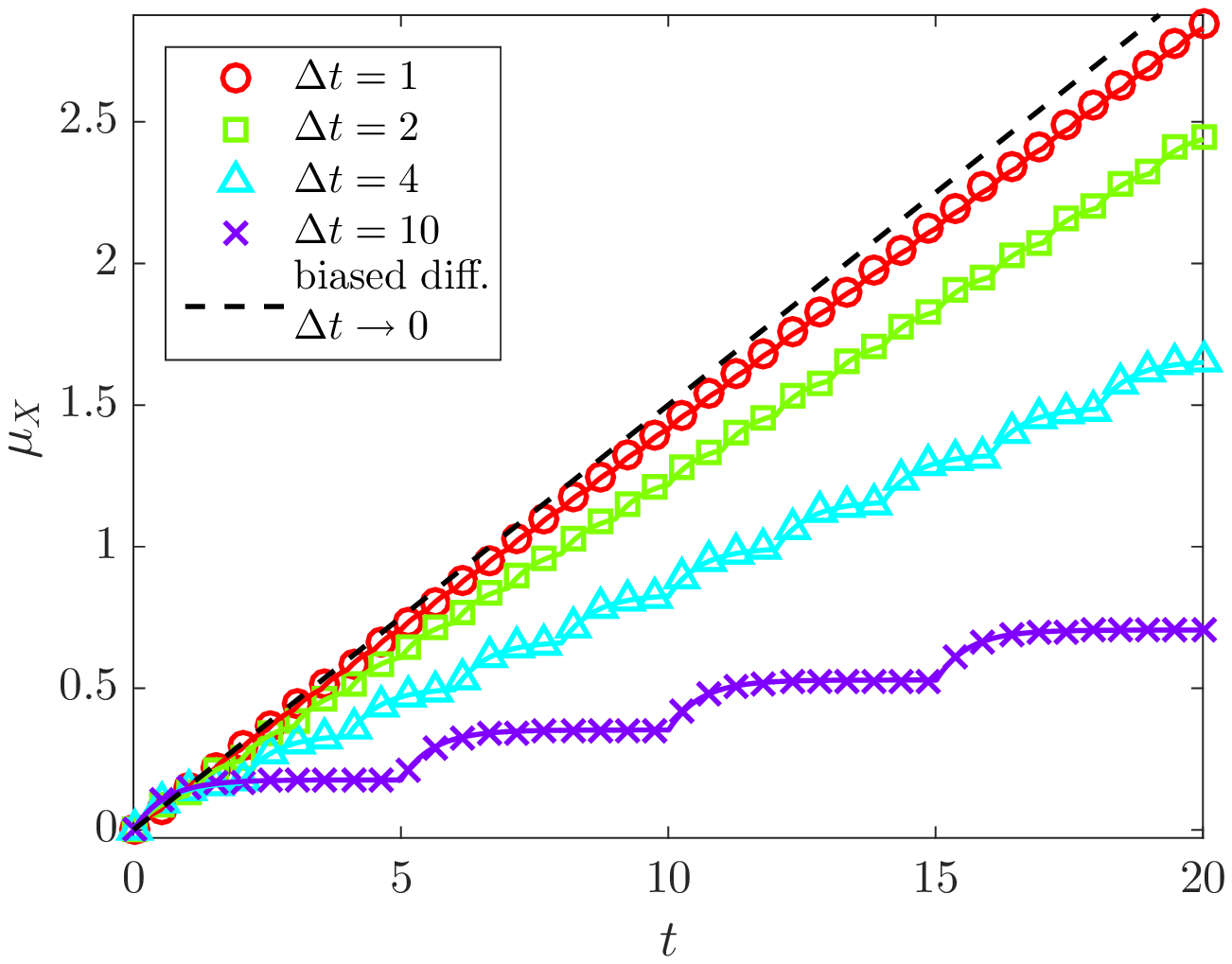}%
\caption{Mean center of mass position $\mu_X(t)$ of the one-dimensional microswimmer for different cycle times $\Delta t$. The shorter the cycle duration, the better the dynamics of the center of mass are described by biased diffusion. The system parameters are $L=3$, $\nu=0.7$, and $T=0.2$. Symbols represent simulations of the Langevin Eqs.~\eqref{eqn_1dLangevin} of the complete dynamics ($N=10^5$ trajectories, time step $dt = 10^{-2}$).}
\label{fig_1dcheckbias}
\end{figure}

The corresponding Langevin equation describes \emph{biased diffusion} (see, e.g., Ref.~\cite{Gardiner2004_523a}):
\begin{align}
\dot{X} =\nu_{\mathrm{eff}} f_{\mathrm{eff}} + \sqrt{2 \nu_{\mathrm{eff}}\, T}\,\xi(t).
\end{align}
This first central finding shows that the complex microswimmer dynamics simplify to biased diffusion of the center of mass in the limit of small cycle times. Figure~\ref{fig_1dcheckbias} shows how the mean value $\mu_X(t)$ approaches the limit of biased diffusion where $\mu(t) =\nu_{\mathrm{eff}} f_{\mathrm{eff}}\,t $ when $\Delta t \rightarrow 0$. Similar results hold for the variance $c_{XX}$.

\section{two-dimensional swimmer}\label{sec:2dswimmer}
We proceed to analyze the model in two dimensions. Here, it has a richer structure as there is an additional rotational diffusion of the swimmer. The particles are at positions $(x_1,y_1)$ and $(x_2,y_2)$, respectively. Their distance is given by $r~=~\sqrt{(x_2-x_1)^2 + (y_2-y_1)^2}$. For the interaction potential we choose
\begin{align}\label{eqn_2dpotential}
V(r;l(t)) = \frac{1}{r} + \frac{1}{2} (r - l(t))^2,
\end{align} 
which now also contains a repulsive term needed to enable smooth rotational diffusion as we shall see later.

The Langevin Eqs.~\eqref{eqn_Langevin} read: 
\begin{subequations}\label{eqn_2dLangevin}
\begin{align}
	\dot{x}_1 &= \nu_1 V' \frac{x_2-x_1}{r} + \sqrt{2 \nu_1 T} \xi_{1}^{(x)}, \\
	\dot{y}_1 &= \nu_1 V'\frac{y_2-y_1}{r} + \sqrt{2 \nu_1 T} \xi_{1}^{(y)}, \\
	\dot{x}_2 &= - \nu_2 V' \frac{x_2-x_1}{r} + \sqrt{2 \nu_2 T} \xi_{2}^{(x)}, \\
	\dot{y}_2 &= - \nu_2 V' \frac{y_2-y_1}{r} + \sqrt{2 \nu_2 T} \xi_{2}^{(y)}.
\end{align}
\end{subequations}
Introducing the angle $\phi := \arctan{\frac{y_2-y_1}{x_2-x_1}}$ and the center of mass coordinates $X~:=~\frac{1}{2}(x_1 + x_2)$ and $Y := \frac{1}{2}(y_1 + y_2)$ we obtain
\begin{subequations}\label{eqn_2dRelativeLangevin}
\begin{align}
	\dot{X} &= \frac{\nu_1 - \nu_2}{2} V' \cos{\phi} +  \sqrt{\frac{\nu_1 T}{2}} \xi_{1}^{(x)} + \sqrt{\frac{\nu_2 T}{2}} \xi_{2}^{(x)}, \\
	\dot{Y} &= \frac{\nu_1 - \nu_2}{2} V' \sin{\phi} + \sqrt{\frac{\nu_1 T}{2}} \xi_{1}^{(y)} + \sqrt{\frac{\nu_2 T}{2}} \xi_{2}^{(y)}, \\	
	\dot{r} &= - (\nu_1 + \nu_2) V' + \sqrt{2 T}\left( \cos{\phi}\, \zeta^{(x)} + \sin{\phi}\,\zeta^{(y)}\right), \\
	\dot{\phi} &= \frac{\sqrt{2T}}{r} \left( \cos{\phi}\, \zeta^{(y)} - \sin{\phi}\,\zeta^{(x)}\right), 
\end{align}
\end{subequations}
where 
\begin{subequations}
\begin{align}
\zeta^{(x)} &= \sqrt{\nu_2} \,\xi_2^{(x)} -\sqrt{\nu_1}  \,\xi_1^{(x)},\\
\zeta^{(y)} &= \sqrt{\nu_2}\,\xi_2^{(y)} -\sqrt{\nu_1}\, \xi_1^{(y)}.
\end{align}
\end{subequations}

If the Langevin Eqs.~\eqref{eqn_2dRelativeLangevin} are interpreted in the Stratonovich sense, the corresponding Fokker-Planck equation for the joint distribution $p = p(X,Y,r,\phi;t)$ reads~\cite{Gardiner2004_436}
\begin{align}
\partial_t p &= \left\{ \mathcal{L}_{r\phi} + \mathcal{L}_{XY}+ (\nu_2-\nu_1)T \left[\cos{\phi} \left( \partial^2_{r X} + \frac{1}{r} \partial^2_{\phi Y} \right) \right.\right.  \nonumber\\ 
&\qquad\qquad\;\qquad+ \left.\left. \sin{\phi} \left( \partial^2_{rY} - \frac{1}{r}\partial^2_{\phi X} \right) \right] \right\}\, p,
\end{align}
with
\begin{align}
\mathcal{L}_{r\phi} &= (\nu_1 + \nu_2) \left[ \partial_r \left( V' - \frac{T}{r} \right) + T\, \partial^2_r + \frac{T}{r^2}\, \partial^2_\phi  \right]
\end{align}
and
\begin{align}
\mathcal{L}_{X Y} &= (\nu_2-\nu_1) \left( \frac{V'}{2} -\frac{T}{r} \right) \left( \cos{\phi}\, \partial_X + \sin{\phi} \partial_Y \,\right)  \nonumber\\
&\qquad\qquad\qquad\qquad + \frac{\nu_1+\nu_2}{4} T \left( \partial^2_X + \partial^2_Y \right).
\end{align}

\subsection{Coarse-graining in the limit of short cycle times}\label{sec:2dCG}
We now investigate the limit $\Delta t \rightarrow 0$. In analogy to Sec.~\ref{sec:1dCG}, we use Eqs.~\eqref{eqn_effectiveFPE},~\eqref{eqn_effectiveGenerator},~and~\eqref{eqn_2dpotential} to obtain an effective Fokker-Planck equation:
\begin{align}\label{eqn_2dLimitFPE}
\partial_t p(r,\phi,X,Y;t) &= \left( \bar{\mathcal{L}}_{r\phi} + \bar{\mathcal{L}}_{XY} \right) p(X,Y,r,\phi;t),
\end{align}
where
\begin{align}
\bar{\mathcal{L}}_{r\phi} &= \left(1+\nu \right) \partial_r \left(-\frac{1}{r^2}-\frac{T}{r}+ r - \frac{L+1}{2}\right) \nonumber\\
&\qquad\qquad\qquad\qquad + (1+\nu) T\,\left(\partial^2_r +\frac{1}{r^2} \partial^2_\phi \right),\\
\bar{\mathcal{L}}_{XY} &= \nu_{\mathrm{eff}} f_{\mathrm{eff}} \left( \cos{\phi}\, \partial_X + \sin{\phi}\, \partial_Y \right)  + \nu_{\mathrm{eff}} T \left( \partial_X^2 +\partial_Y^2 \right).
\end{align}
The effective mobility $\nu_{\mathrm{eff}}$ and constant force $f_{\mathrm{eff}}$ are again given by Eqs.~\eqref{eqn_effMobility}~and~\eqref{eqn_effForce}, respectively.

Integrating Eq.~\eqref{eqn_2dLimitFPE} over $X$, $Y$, and $\phi$, we obtain a Fokker-Planck equation for the marginal distribution $p_r = p_r(r;t)$ of the relative coordinate:
\begin{align}\label{eqn_CG2d_r}
\frac{1}{1+\nu}\partial_t p_r = \left[ \partial_r \left(-\frac{1}{r^2}-\frac{T}{r}+ r - \frac{L+1}{2}\right) +  T\partial^2_r \right] p_r.
\end{align}
Its solution for long times $t$ yields the steady-state distribution of $r$:
\begin{align}
p_r^{\mathrm{st}}(r) = \frac{r}{Z} \,\exp{\left[- \frac{1}{T}V\left(r;\frac{L+1}{2}\right)  \right]},
\end{align}
where $Z$ ensures normalization and $V(r;l)$ is given by Eq.~\eqref{eqn_2dpotential}.

As before, we assume that the relative coordinate has reached its periodic steady state. Thus, with the ansatz $p(X,Y,r,\phi;t) = p(X,Y,\phi;t)\,p_r^{\mathrm{st}}(r)$ and using Eq.~\eqref{eqn_CG2d_r} we obtain the Fokker-Planck equation for the center of mass movement and the direction of the swimmer:
\begin{align} \label{eqn_2dEffectiveFPE}
\partial_t p(X,Y,\phi;t) = \Big[-\nu_{\mathrm{eff}}\,f_{\mathrm{eff}}\left(  \cos{\phi}\, \partial_X + \sin{\phi}\, \partial_Y \right) \nonumber\\
 +  \nu_{\mathrm{eff}}T\,\left( \partial^2_X +\partial^2_Y \right) + \partial^2_\phi \, D_\phi \Big] p(X,Y,\phi;t).
\end{align}
The directional diffusion $D_\phi$ is given by
\begin{align}\label{eqn_2dDiffsuionCoeffRot}
D_\phi = 4 \nu_{\mathrm{eff}} T\int_0^\infty dr\, \frac{p_r^{\mathrm{st}}(r)}{r^2} = \mathrm{const}.
\end{align}

Here, we see that an additional repulsive term in the potential in Eq.~\eqref{eqn_2dpotential} is needed: Otherwise, the integral in Eq.~\eqref{eqn_2dDiffsuionCoeffRot} diverges at the lower limit and the rotational dynamics cannot be described by simple diffusion.

Thus, for short cycle times, the center of mass movement is given by \emph{active Brownian motion}~\cite{Bechinger2016}:
\begin{subequations}\label{eqn_ABM}
\begin{align}
\dot{X} &=\nu_{\mathrm{eff}} f_{\mathrm{eff}} \, \cos{\phi} + \sqrt{2 \nu_{\mathrm{eff}}\, T}\,\xi^{(X)}(t),\\
\dot{Y} &=\nu_{\mathrm{eff}} f_{\mathrm{eff}} \, \sin{\phi} + \sqrt{2 \nu_{\mathrm{eff}}\, T}\,\xi^{(Y)}(t),\\
\dot{\phi} &= \sqrt{2 D_\phi}\, \xi^{(\phi)}(t),
\end{align}
\end{subequations}
which is a two-dimensional generalization of biased diffusion. This constitutes our second main finding.

\begin{figure}[ht]
\includegraphics[width=1\linewidth]{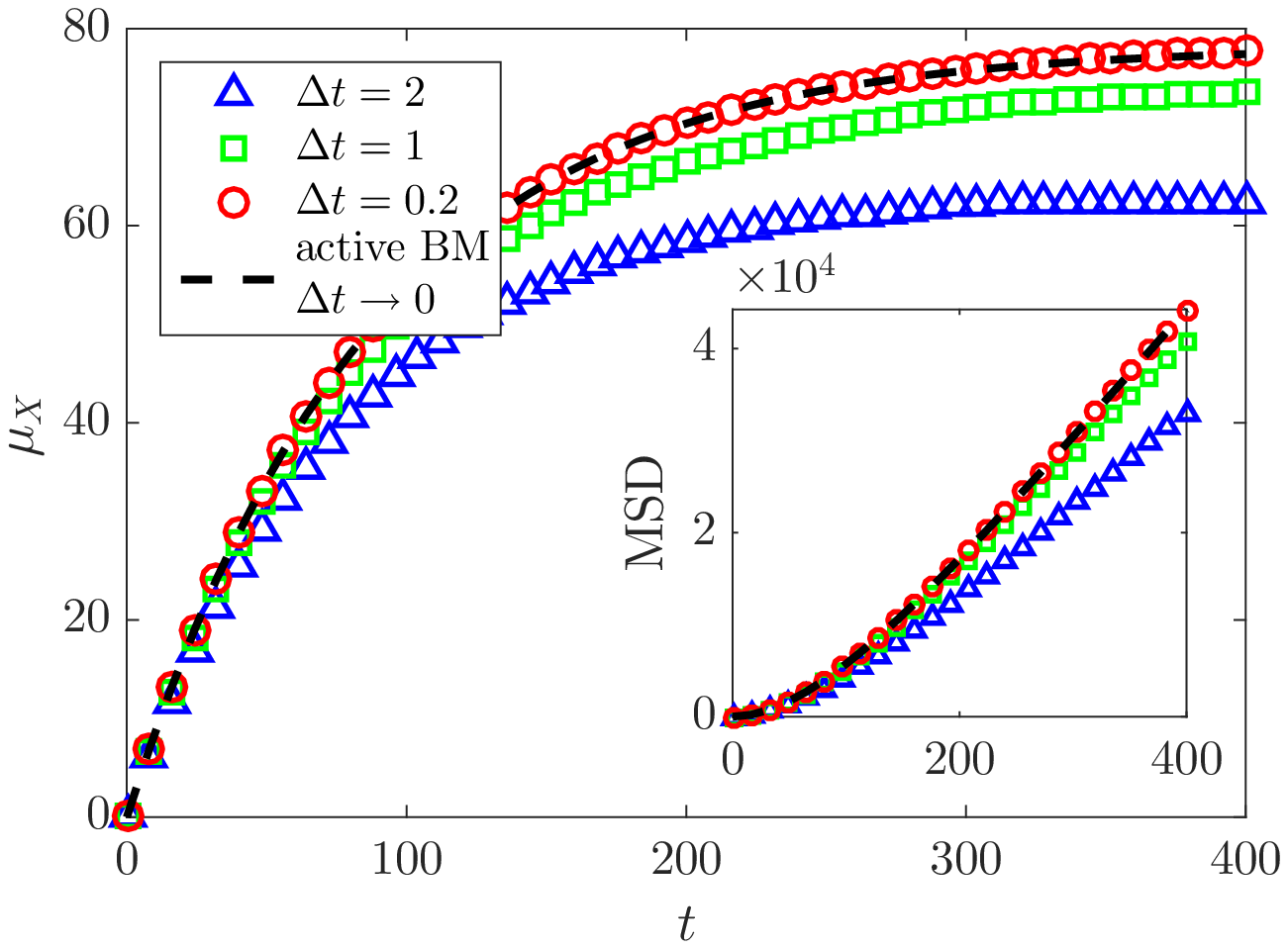}
\caption{Mean center of mass position $\mu_X(t)$ and mean-squared displacement (inset) of the two-dimensional model for different cycle times $\Delta t$. The shorter the cycle duration, the better the dynamics of the center of mass are described by active Brownian motion. The system parameters are $L=5$, $\nu=0.1$, and $T=0.1$. Symbols represent simulations of the Langevin Eqs.~\eqref{eqn_2dLangevin} of the complete process ($N=10^5$ trajectories, time step $dt = 10^{-2}$).}
\label{fig_2dcheckABM}
\end{figure}

If the process described by Eqs.~\eqref{eqn_ABM} is started from $X=Y=\phi=0$, the time-dependent mean $\mu_X(t)$ and mean-squared displacement $\mathrm{MSD}(t)$ are given by:
\begin{subequations}
\begin{align}
\mu_X(t) &= \frac{\nu_{\mathrm{eff}} f_{\mathrm{eff}}}{D_\phi} \left( 1- e^{ -D_\phi t } \right),\\
\mathrm{MSD}(t) &= \left( 4 \nu_{\mathrm{eff}} T + \frac{2 \nu^2_{\mathrm{eff}}f_{\mathrm{eff}}^2}{D_\phi} \right)t - \frac{2\nu^2_{\mathrm{eff}}f_{\mathrm{eff}}^2}{D_\phi^2}\left( 1- e^{ -D_\phi t } \right).
\end{align}
\end{subequations}
Figure~\ref{fig_2dcheckABM} shows how the complete process approaches this limiting case as $\Delta t \rightarrow 0$.

We therefore see that the center of mass movement of our microswimmer model is described by biased diffusion in one dimension and active Brownian motion in two dimensions when the cycle times become short.

\section{Comparison of dissipation rates}\label{sec:compareEnergies}
To sustain its motion, any microswimmer must convert energy into heat that is dissipated into the surrounding medium. In this section, we calculate the rate of energy dissipation for the real microswimmer model and compare it to the dissipation rate that is inferred from the coarse-grained active Brownian motion.

For the one-dimensional swimmer, the complete dissipation per cycle $\Delta Q$ can be easily calculated by realizing that the average potential energy is periodic. Using the first law~\cite{Sekimoto1998} and realizing that the work done on the system only has contributions from the abrupt changes in the interaction potential we find
\begin{align}
\Delta Q &= \Delta W\nonumber\\
&= \bigg\langle V\left(r\left(\frac{\Delta t}{2} \right), 1\right) - V\left(r\left(\frac{\Delta t}{2} \right), L\right)  \nonumber\\
 &\qquad\qquad+ V\left(r(0), L\right) - V\left(r\left(0 \right), 1\right)   \bigg\rangle\nonumber\\
 &= (L-1)\,\left[ \mu_r\left(\frac{\Delta t}{2}\right) - \mu_r(0)\right]\label{eqn_DeltaQ_depon_r}\\
 &= (L-1)^2 \, \tanh\left( \frac{1+\nu}{4} \Delta t \right),\label{eqn_DeltaQ_res}
\end{align}
where we have used Eq.~\eqref{eqn_1d_solmur}.

For small cycle times, the rate $\dot{Q}$ of energy dissipation thus reads
\begin{align}\label{eqn_realHeatLimit}
\dot{Q} = \lim\limits_{\Delta t \rightarrow 0 } \frac{\Delta Q}{\Delta t} = \frac{(L-1)^2\,(1+\nu)}{4}.
\end{align}

In contrast, the energy dissipation rate assigned to the effective process reads, following Sekimoto's definition~\cite{Sekimoto1998},
\begin{align}\label{eqn_effectiveHeatLimit}
\dot{Q}_{\mathrm{eff}} = \left\langle \dot{X}\, f_{\mathrm{eff}} \right\rangle = \nu_{\mathrm{eff}}f_{\mathrm{eff}}^2 = \frac{(L-1)^2\, (1-\nu)^2}{4\,(1+\nu)},
\end{align} 
where we used Eqs.~\eqref{eqn_effMobility}~and~\eqref{eqn_effForce}. The ratio of these dissipation rates is given by
\begin{align}\label{eqn_ratioHeat}
\frac{\dot{Q}_{\mathrm{eff}}}{\dot{Q}} = \frac{(1-\nu)^2}{(1+\nu)^2} \leq 1.
\end{align}

\begin{figure}[ht]
\includegraphics[width=1\linewidth]{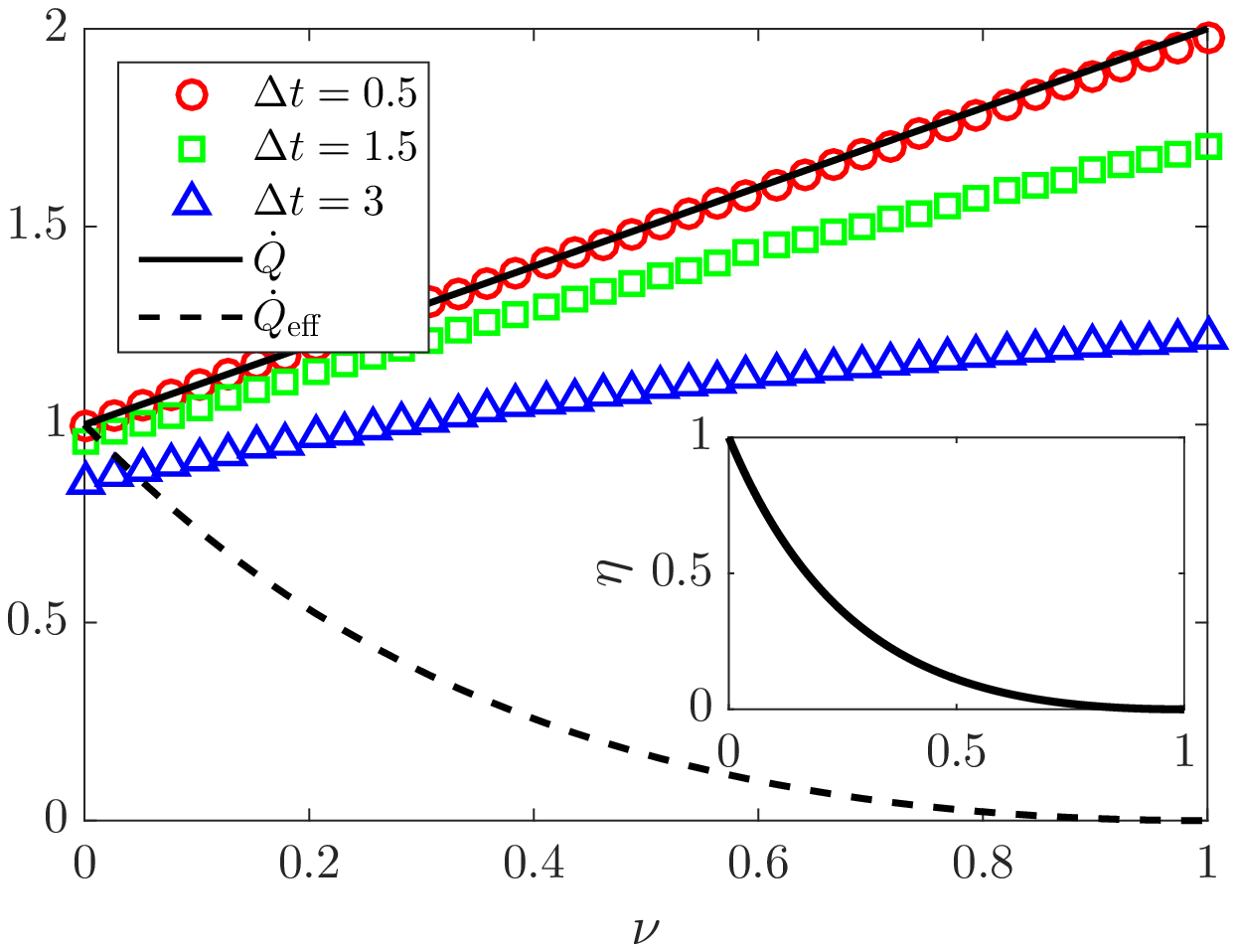}
\caption{Comparison of energy dissipation rates. Symbols represent average dissipation rates obtained from applying Sekimoto's definition of heat to trajectories obtained from simulations of the complete process ($N=10^4$ trajectories, time step $dt = 10^{-2}$). The short cycle limit is shown as a solid line. Additionally the effective dissipation of the biased diffusion process is shown (dashed line). For the simulation the system parameters are $L=3$ and $T=0.2$. Inset: efficiency $\eta$ of the swimmer.}
\label{fig_1ddissipation}
\end{figure}

Figure~\ref{fig_1ddissipation} shows how the complete energy dissipation rate approaches the limiting rate in Eq.~\eqref{eqn_realHeatLimit}. The simulation results are obtained by applying Sekimoto's definition to the complete system, i.e., calculating force times velocity for both particles. For comparison the effective dissipation in Eq.~\eqref{eqn_effectiveHeatLimit} is also shown.

From Eq.~\eqref{eqn_ratioHeat} as well as from Fig.~\ref{fig_1ddissipation}, we infer that the effective dissipation rate always underestimates the complete dissipation rate. Interestingly, with increasing $\nu$ the total dissipation grows while the effective dissipation decreases. There can even be an extreme discrepancy between them as the effective dissipation vanishes when the complete dissipation is maximal. They agree only when $\nu=0$, i.e., where the particle with the low mobility cannot move at all.

To better understand this issue, let us define an efficiency $\eta$ of the swimming mechanism by taking the ratio of the average energy $\Delta Q_X$ dissipated in one cycle by moving the center of mass to the complete average dissipation: 
\begin{align}\label{eqn_efficiency}
\eta :=  \frac{\Delta Q_X}{\Delta Q}.
\end{align}
The heat $\Delta Q_X$ is given by
\begin{align}\label{eqn_heatCOM}
\Delta Q_X := \int_0^{\Delta t} dt\,\left\langle \dot{X}\, f_X(r;t) \right\rangle,
\end{align}
where $f_X(r;t)$ is the force on the center of mass. According to Eqs.~\eqref{eqn_1dRelativeLangevin}~and~\eqref{eqn_effMobility}, this is given by
\begin{align}\label{eqn_forceCOM}
f_X(r;t) = \frac{4}{1+\nu }\frac{\nu_1-\nu_2}{2} \, V'(r;l(t)).
\end{align}
As we show in Appendix~\ref{sec:A_heatCOM}, this dissipated energy reads
\begin{align}\label{eqn_resultHeatCOM}
\Delta Q_X = \frac{(1-\nu)^2}{(1+\nu)^2}(L-1)^2\, \tanh\left( \frac{1+\nu}{4} \Delta t \right).
\end{align}
For $\Delta t \rightarrow 0$, we recover the dissipation rate of the effective process.

Therefore, with Eqs.~\eqref{eqn_DeltaQ_res},~\eqref{eqn_ratioHeat},~\eqref{eqn_efficiency},~and~\eqref{eqn_resultHeatCOM}, the efficiency is the same as the ratio of the effective dissipation to the complete dissipation:
\begin{align}\label{eqn_ratioHeatCOMtoReal}
\eta = \frac{(1-\nu)^2}{(1+\nu)^2} = \frac{\dot{Q}_{\mathrm{eff}}}{\dot{Q}}.
\end{align}

This efficiency is plotted in the inset of Fig.~\ref{fig_1ddissipation}. It is monotonously decreasing from maximum to vanishing efficiency with increasing $\nu$.

We also investigated the dissipation rates for the two-dimensional model with similar results. The calculations can only be carried out numerically as outlined in Appendix~\ref{sec:B_2ddiss}.

\section{Discussion}\label{sec:discussion}
The dissipation assigned to the active Brownian motion approximation underestimates the complete dissipation occurring in the full model. This fact matches previous results by Esposito~\cite{Esposito2012} showing that a coarse-grained average entropy production underestimates the true average entropy production. This is a fairly general result but the magnitude of the discrepancy is left open.

However,  with our specific model at hand, we can quantify the difference between the observed and the complete dissipation. Depending on the parameter configuration, it can be extremely small or large: For $\nu \rightarrow 0$, the complete dissipation is perfectly captured by the observed dissipation while for $\nu \rightarrow 1$, it is grossly underestimated.

That is because when observing the center of mass movement, one only glimpses at traces of the total dissipation. This total dissipation depends solely on the relative coordinate $r$ as can be seen in Eq.~\eqref{eqn_DeltaQ_depon_r}. Equations~\eqref{eqn_resultHeatCOM}~and~\eqref{eqn_ratioHeatCOMtoReal} imply that only part of this dissipation results in forward propulsion of the center of mass.

Knowing the changes in the center of mass position merely gives a part of the information needed to infer the complete dissipation. Only when the particles are alternately immobile ($\nu \rightarrow 0$) is the total dissipation captured by the center of mass displacement. In that case, changes in the relative coordinate are strictly proportional to translations of the center of mass. For $\nu \rightarrow 1$ the mobilities of both particles are almost equal and the microswimmer wastes energy in expanding and contracting while achieving minimal propulsion.

This justifies the definition of an efficiency of a microswimmer as the ratio of dissipated energy utilized for useful forward propulsion to the total dissipation. This efficiency is maximized for active Brownian motion as all energy is dissipated in forward propulsion. It measures the deviations of more complicated swimming strategies from this optimum. This can be seen in our model as well: For $\nu=0$, our swimmer invests all dissipation in forward propulsion. Consequently, it is maximally efficient. 

To derive these results, we showed that the center of mass movement of a microswimmer with periodic driving can be mapped onto active Brownian motion when the cycle time becomes short. This is especially relevant for experiments as the swimming dynamics are often fast and spacial imaging resolution is usually limited, enabling only a tracking of the body of the swimmer.

Note that the additional repulsive term in the potential in Eq.~\eqref{eqn_2dpotential} needed to enable smooth rotational diffusion of the two-dimensional model is only an issue in theoretical modeling. In reality, there is a \emph{hard core repulsion} keeping the particles at least two radii apart.

We need to point out that while active Brownian motion [Eq.~\eqref{eqn_ABM}] correctly describes the ensemble distribution of the coordinates $X$, $Y$, and $\phi$, on the level of individual trajectories the description is not correct. Particularly, the $\phi$-process is not Markovian. This is a consequence of the coarse-graining we performed by integrating out the $r$-variable to arrive at Eq.~\eqref{eqn_2dEffectiveFPE}. It is known that coarse-graining preserves the ensemble distribution of visible variables but it does not yield the correct description of the trajectory probabilites~\cite{Kahlen2018}. This effect does not arise in the one-dimensional model as there is no coupling between $r$ and $X$ after taking the limit $\Delta t \rightarrow 0$ [cf.~Eq.~\eqref{eqn_effectiveFPE}].

Our results show that active Brownian motion can be a good approximation for microswimmer dynamics. The findings can help to gauge the quality of this approximation for the energetics of microswimmers, especially if they have additional degrees of freedom which are not correctly resolved.

\section{Conclusion}\label{sec:conclusion}
We analyzed the energetics of a microswimmer consisting of a system of two coupled Brownian particles able to generate self propulsion. For fast internal dynamics, the center of mass movement obeys biased diffusion in one dimension and active Brownian motion in two dimensions. We quantified the difference between the actual dissipation and the effective dissipation captured by active Brownian motion and showed that there can be a large discrepancy between these descriptions even though the observed dynamics are the same. This is due to the fact that some parts of the system where dissipation occurs are not observed. We introduced a swimming efficiency that captures how much of the dissipation is used in actual propulsion.

\begin{acknowledgments}
We thank Andreas Engel for valuable discussions and critically reading the paper.

Both authors contributed equally to this paper.
\end{acknowledgments}

\appendix
\begin{widetext}
\section{Dissipation by the center of mass} \label{sec:A_heatCOM}
Following Eqs.~\eqref{eqn_heatCOM}~and~\eqref{eqn_forceCOM} and using the definitions of center of mass and relative coordinates, the average energy dissipated by the center of mass during one cycle is given by
\begin{align}
\Delta Q_X :=  \int\limits_0^{\Delta t}dt\left\langle (\dot{x}_1+\dot{x}_2)\, \frac{\nu_1-\nu_2}{1+\nu}\,(x_2-x_1-l) \right\rangle,
\end{align}
which can be simplified to~\cite{Seifert2005}
\begin{align}
\Delta Q_X :=  \int\limits_0^{\Delta t} dt \iint dx dy\, (j_1+j_2)\, \frac{\nu_1-\nu_2}{1+\nu}\,(x_2-x_1-l),
\end{align}
where 
\begin{align}
j_1 &= \left[\nu_1 (x_2-x_1-l) - \nu_1 T\partial_{x_1}\right] \, p(x_1,x_2;t)\\
j_2 &= \left[- \nu_2 (x_2-x_1-l) - \nu_2 T\partial_{x_2}\right] \, p(x_1,x_2;t)
\end{align}
are the probability currents of the Fokker-Planck equation corresponding to the Langevin Eqs.~\eqref{eqn_1dLangevin}. The joint probability $p(x_1,x_2;t)$ can be calculated from $p(r,X;t)$ by transformation of variables. We then obtain
\begin{align}
\Delta Q_X = \frac{(1-\nu)^2}{(1+\nu)}\left[\int\limits_0^{\Delta t/2} dt  \left(\mu_r-L\right)^2 +\int\limits_{\Delta t /2}^{\Delta t} dt  \left(\mu_r-1\right)^2 \right],
\end{align}
and the result presented in Eq.~\eqref{eqn_resultHeatCOM} follows with Eq.~\eqref{eqn_1d_solmur}.

\section{Dissipation in two dimensions} \label{sec:B_2ddiss}
We calculate the dissipation of the two dimensional microswimmer. The dissipation per cycle $\Delta Q$ is given by
\begin{align}\label{eqn_2ddiss}
	\Delta Q = (L-1)\,\left[ \mu_r\left(\frac{\Delta t}{2}\right) - \mu_r(0)\right],
\end{align}
analogously to Eq.~\eqref{eqn_DeltaQ_depon_r}.

The mean values in the above equation cannot be calculated directly. Instead, we obtain an approximation valid for small $\Delta t$. 

First, from Eqs.~\eqref{eqn_2dLangevin} we obtain the Fokker-Planck equation for the variables $r_x~:=~x_2-x_1$ and $r_y~:=~y_2~-~y_1$. From this, we find the infinitesimal propagator~\cite{Risken1996}. A transformation of variables to $r$ and $\phi$ such that $r_x = r\cos{\phi}$ and $r_y = r\sin{\phi}$, subsequent integration over $\phi$, and an expansion in the exponent up to terms of order $dt$ yields
\begin{align}
p(r', t+dt | r, t) &= \sqrt{\frac{r'/r}{4 \pi \tilde{T} dt}} \exp\left[ -\frac{(r' - r + dt \tilde{V}')^2}{4 \tilde{T} dt} \right] \,\exp\left[ \frac{\tilde{T} dt}{4 r r'} + dt \frac{\tilde{V}'}{2 r} \right],
\end{align}
where $\tilde{T} = (1+\nu) T$ and $\tilde{V}'=(1+\nu) V'$.

Thus, for small cycle times $\Delta t$, the propagator for one cycle reads 
\begin{align}\label{eqn_PropSmallCycle}
	p(r'', \Delta t | r, 0) = \int\limits_0^{\infty} dr' p\left(r'', \Delta t \big| r', \frac{\Delta t}{2}\right) p\left(r', \frac{\Delta t}{2} \big| r, 0\right) .
\end{align}

The distribution $p_r(r, 0)$ is numerically obtained by discretizing the propagator of one cycle in Eq.~\eqref{eqn_PropSmallCycle} in $r$ and $r''$. The eigenvector to the eigenvalue $1$ is the distribution $p_r(r, 0)$ from which we obtain the average $\mu_r(0)$. 
The second average $\mu_r\left(\frac{\Delta t}{2}\right)$ then follows from 
\begin{align}
	p_r\left(r, \frac{\Delta t}{2}\right) = \int_0^{\infty} dr' p\left(r, \frac{\Delta t}{2} \big| r', 0\right) p_r(r', 0) .
\end{align}
A comparison of the numerical results with a simulation is given in Fig.~\ref{fig_radialaverage}.

\begin{figure}[ht]
 \includegraphics[width=0.5\linewidth]{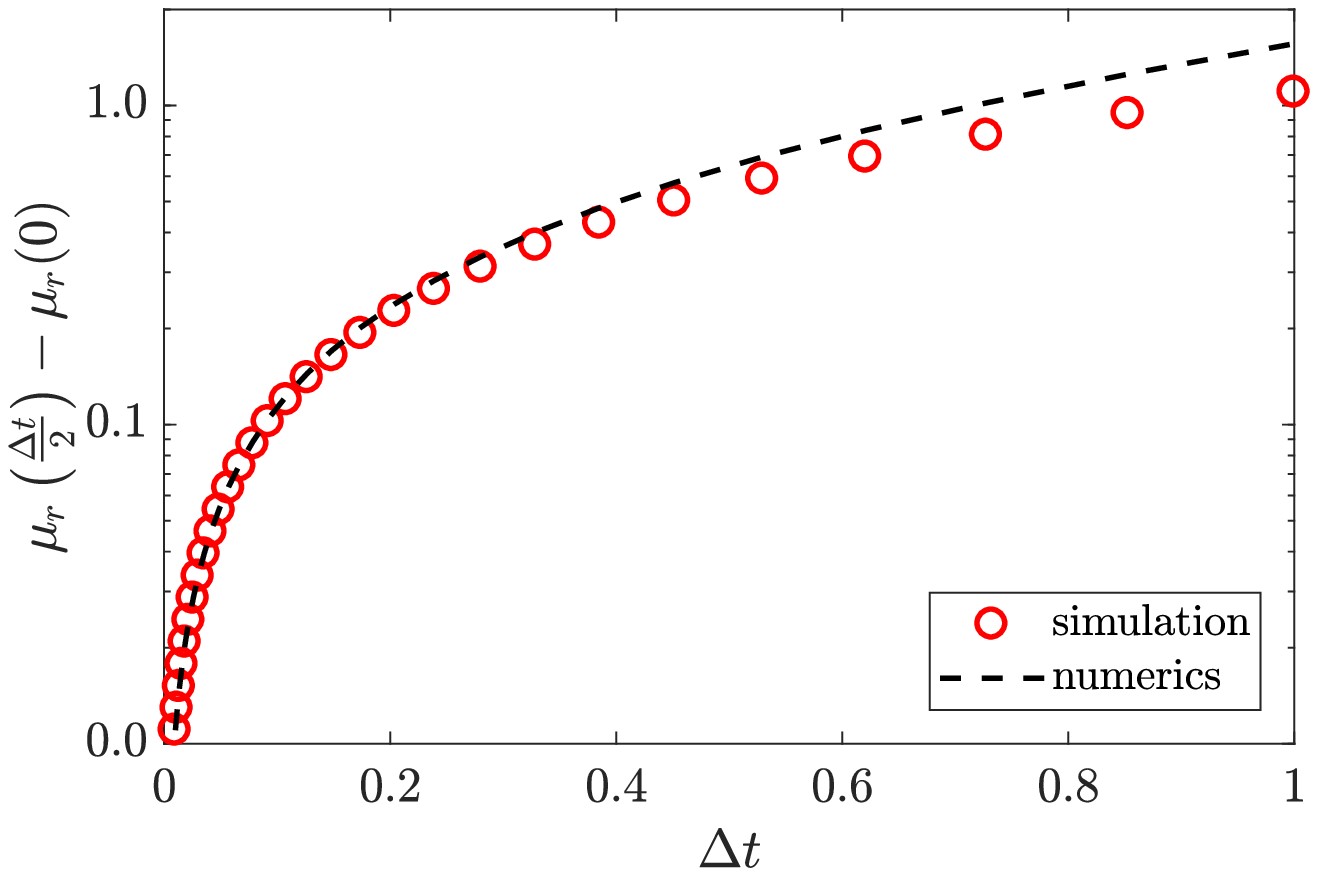}
 \caption{Mean values for the relative coordinate of the two-dimensional swimmer and comparison with simulations ($ 10^4 $ trajectories, time step $dt=10^{-2}$) of the Langevin Eqs.~\eqref{eqn_2dLangevin} for system parameters $L=5$, $T = 0.1$, and $\nu = 0.1$.}
 \label{fig_radialaverage}
\end{figure}

These mean values can be inserted into Eq.~\eqref{eqn_2ddiss}, yielding the complete dissipation for small $\Delta t$. The dissipation rate of the active Brownian motion is again given by Eq.~\eqref{eqn_effectiveHeatLimit}. A comparison of the two dissipation rates yields qualitatively similar results to those depicted in Fig.~\ref{fig_1ddissipation}.
\end{widetext}

\bibliography{references}

\end{document}